\documentclass[10pt,nolongbibliography,twocolumn,superscriptaddress,prb,showpacs,showkeys]{revtex4-2}

\usepackage{graphicx} % Required for inserting images
\usepackage{amsmath}
\usepackage{hyperref}
\usepackage[caption=false]{subfig}
\usepackage{float}
\usepackage{pgffor}

\begin{document}
\title{Core-hole induced misalignment between Van-Hove singularities and \emph{K}-edge fine structure in carbon nanotubes}
\author{Martin Unzog}
\email{martin.unzog@univie.ac.at}
\affiliation{%
University of Vienna, Faculty of Physics \& Computational Materials Physics \& Vienna Doctoral School in Physics,  Boltzmanngasse 5, 1090 Vienna, Austria
}%

\author{Alexey Tal}
\email{alexey.tal@vasp.at}
\affiliation{%
VASP Software GmbH, Berggasse 21/14, A-1090, Vienna, Austria
}%

\author{Pedro Melo}
\email{pedro.melo@vasp.at}
\affiliation{%
VASP Software GmbH, Berggasse 21/14, A-1090, Vienna, Austria
}%

\author{Ryosuke Senga}
\email{ryosuke-senga@aist.go.jp}
\affiliation{%
Nano-Materials Research Institute, National Institute of Advanced Industrial Science and Technology (AIST), Tsukuba 305-8565, Japan
}%

\author{Kazu Suenaga}
\email{suenaga-kazu@sanken.osaka-u.jp}
\affiliation{%
The Institute of Scientific and Industrial Research (ISIR), Osaka University; Mihogaoka 8-1, Ibaraki, Osaka 567-0047, Japan
}%

\author{Thomas Pichler}
\email{thomas.pichler@univie.ac.at}
\affiliation{%
Faculty of Physics, University of Vienna, Strudlhofgasse 4, A-1090 Vienna, Austria
}%

\author{Georg Kresse}%
\email{georg.kresse@univie.ac.at}
\affiliation{%
University of Vienna, Faculty of Physics, Computational Materials Physics, Kolingasse 14-16, 1090 Vienna, Austria
 }%
\begin{abstract}

We investigate the relationship between the \emph{K}-edge fine structure of isolated single-wall carbon nanotubes (SWCNTs) and the Van Hove singularities (VHSs) in the conduction band density of states. To this end, we model X-ray absorption spectra of SWCNTs using the final-state approximation and the Bethe-Salpeter equation (BSE) method. Both methods can reproduce the experimental fine structure, where the BSE results improve on peak positions and amplitude rations compared to the final-state approximation. When the fine structure in the modeled spectra is related to the VHSs significant differences are found. We suggest that these differences arise due to modifications of the core exciton wavefunctions induced by the confinement along the circumference. Additionally, we analyze the character of core excitons in SWCNTs and find that the first bright excitons are Frenkel excitons while higher-lying excitons are charge resonance states. Finally, we suggest that the qualitative picture based on VHSs in the density of states holds when there is a large energy gap between successive VHSs.

\end{abstract}
\maketitle

\section{Introduction}
%\openup 1em

Single-wall carbon nanotubes (SWCNTs) are cylindrical sheets of graphene \cite{Iijima1991} with unique mechanical, optical, and transport properties that make them interesting for current and future device applications \cite{Harris2009,Schnorr2010,Tan2012,Volder2013,Park2013}. Characteristic of the one-dimensional nature of SWCNTs are the Van Hove singularities (VHSs) visible in their electronic ground-state density of states (DOS) \cite{Saito1998}. These VHSs play an important role in discussing the transport and optical properties of SWCNTs \cite{Charlier2007,Carlson2008,Soavi2016} and their applications in opto- or microelectronics, see e.g. \cite{Liu1999,Nanot2012,Scarselli2012,Zhang2017,Qiu2019,Joo2019}. 

X-ray absorption spectroscopy (XAS) was used to study the unoccupied DOS of metallicity-sorted \cite{Kramberger2007}, diameter-sorted \cite{Ayala2009}, and mainly (6,5)/(6,4)\cite{Blauwe2010} SWCNTs. These studies have shown a fine structure in the absorption edges. Recently, Senga \emph{et al.}\ \cite{Senga2016} used electron energy loss spectroscopy (EELS) to measure \emph{K}-edges of isolated SWCNTs and found that different chiralities can be distinguished by their unique fine structure in the absorption edges. Kramberger \emph{et al.}\ \cite{Kramberger2007} first used the VHSs to interpret the fine structure in the \emph{K}-edge. This relationship was demonstrated by superimposing a suitably shifted conduction band DOS on the absorption spectrum and associating the first (second, etc.) peak in the fine structure with the first (second, etc.) VHS. This approach was used to explain results on a single (6,5) SWCNT \cite{Senga2016} where it was possible to assign the peaks in the fine structure to the individual VHSs. However, while it was possible to assign each peak in the fine structure to a VHS, a misalignment was visible. In these studies \cite{Kramberger2007,Ayala2009,Blauwe2010,Senga2016} the misalignment was attributed to core-hole effects, \emph{i.e.}\ effects induced by creating the core-hole and, in particular, the formation of excitons. These effects are missing in an independent picture based on the ground-state DOS.

To account for core-hole effects previous studies \cite{Titantah2004,Wessely2006,Gao2009,Mowbray2011} have used methods based on the final-state approximation  \cite{Barth1982,Rehr2000}. These methods remove an electron from the inner shells and add it to the lowest conduction band. A supercell is needed to reduce spurious interactions between excited atoms. The electronic ground-state is then determined self-consistently using Kohn-Sham density functional theory (DFT) \cite{Hohenberg1964,Kohn1965a}. Finally, the spectrum is calculated by Fermi's golden rule using the orbitals determined in the presence of the core-hole. 

When Refs.\ \cite{Titantah2004,Wessely2006,Gao2009,Mowbray2011} were published, experiments were performed on bundles of SWCNTs or diameter-sorted SWCNTs. The modeled spectra had to be averaged and only tentative conclusions could be drawn from experiments: Titantah \emph{et al.}\ \cite{Titantah2004} demonstrated that core-hole effects must be considered to recreate the shape of the experimental spectra. Wessely \emph{et al.}\ \cite{Wessely2006} extended the final-state method to include dynamical effects by using Mahan-Nozières-deDominicis theory \cite{Mahan1967,Nozieres1958}. They concluded that the position of the absorption edge is significantly affected by chirality, while the energy separation between $\pi^*$ and $\sigma^*$ peaks varies for metallic and semiconducting SWCNTs. Mowbray \emph{et al.}\ \cite{Mowbray2011} concluded that the fine structure arises from localized molecular states perpendicular to the tube axis. Finally, Gao \emph{et al.}\ \cite{Gao2009} showed that core-hole effects should be considered to correctly assign peaks in the diameter-sorted spectra to semiconducting or metallic SWCNTs. Gao \emph{et al.}\ \cite{Gao2009} also discussed the spectra's diameter- and chirality dependence by equating the energy differences between the peaks in the fine structure with those of the VHSs. By deriving the VHSs from a zone-folding dispersion relation it is assumed that core-hole effects are negligible. So far it has not been possible to compare simulated and experimental spectra of isolated SWCNTs.

While the final-state approximation yields reliable spectra it also has some drawbacks. It is based on ground-state theory while removing a core-electron creates an excited state. Self-consistently solving the Kohn-Sham equations including the additional core-hole can reproduce some many-body effects, such as the screening by valence electrons or the electron-hole attraction. In contrast, many-body perturbation theory (MBPT) methods are derived for excited states and therefore treat both effects explicitly \cite{Mattuck1992,Fetter2012}. For modeling spectra using MBPT, the state of the art is the $GW$ + Bethe-Salpeter equation method ($GW$+BSE) \cite{Onida2002,Martin2016}. In this approach, the screened Coulomb attraction between electrons and holes is approximated using the random-phase approximation (RPA) \cite{Bohm1951,Ren2012}. Several implementations apply the $GW$+BSE method to X-ray spectra \cite{Olovsson2009,Laskowski2010,Vinson2011,Vinson2022}. Recently, Unzog \emph{et al.}\ \cite{Unzog2022} presented an X-ray BSE implementation within the VASP code. They showed that in bulk carbon systems (diamond, graphite) $GW$+BSE can resolve features missing in spectra based on the final-state approximation. To the best of our knowledge, $GW$+BSE has not yet been applied to study XAS spectra of SWCNTs. However, using $GW$+BSE, Spataru \emph{et al.}\ \cite{Spataru2004} obtained excellent results for optical spectra of small-diameter SWCNTs. 

In this paper, we present X-ray spectra of isolated SWCNTs using the final-state approximation and the $GW$+BSE method. We compare our results to the experimental spectra measured for individual SWCNTs by Senga \emph{et al.}\ \cite{Senga2016}. We will show that the VHSs rarely align with the peaks at the absorption edge. We suggest that these differences arise from effects induced by the confinement along the circumference which affect weakly bound excitons. This will be shown by studying the excitonic properties of a graphene supercell that reproduces the same periodic boundary conditions as the (6,5) SWCNT. 

We first summarize our methods in Sec.\ \ref{sec:methods}. The results are presented in Sec.\ \ref{sec:results}. In Sec.\ \ref{sec:spectra}, experiments are compared to modeled spectra. In Sec.\ \ref{sec:fatband}, we discuss in detail the absorption edge of the (6,5) SWCNT. We discuss the excitonic properties of the graphene supercell in Sec.\ \ref{sec:periodicity}. A discussion follows in \ref{sec:discussion}, and the paper is concluded in Sec.\ \ref{sec:conclusion}. 

\section{Methods}\label{sec:methods}

\begin{table}
    \centering
    \caption{Table listing radii and the BSE parameters used for modeling \emph{K}-edge spectra of semiconducting SWCNTs.}
\begin{tabular}{l c c c c c c}
    \hline\hline
    Chirality & (6,5) & (9,1) & (8,3) & (8,4) & (11,0) & (10,2)\\ 
    \hline
    Radius (Å) & 3.5 & 3.8 & 3.9 & 4.2 & 4.3 & 4.4 \\
    Unit cell length (Å) & 41.1& 41.1\textbf{} & 42.4 & 11.4 & 4.3 & 24.0\\      
    $1\times1\times n$ $\mathbf{k}$-points & 2 & 2 & 2 & 8 & 18 & 4\\
    Bands used for $W_0$ & 2000 & 2100 & 1500 & 900 & 260 & 1400 \\ 
    Complex frequencies & 24 & 20 & 14 & 24 & 20 & 20\\ 
    Unoccupied bands & 1200 & 1400 & 1200 & 405 & 170 & 1000\\ 
    \hline\hline
\end{tabular}
    
    \label{tab:bseparam}
\end{table}

The plane-wave code VASP \cite{kresse1996,kresse1999} was used to calculate the spectra. The SWCNTs were placed into tetragonal cells with an inter-tube separation of at least 6 Å. The geometries were then relaxed using the Perdew, Burke, Ernzerhof (PBE) functional \cite{Perdew1996} until forces converged below 0.01 eV/Å. 

The orbitals were calculated using the hybrid functional PBE0($\alpha=0.24$) \cite{Perdew1996a,Ernzerhof1999,Adamo1999}. Using this fraction of exact exchange $\alpha=0.24$, the $G_0W_0$ bandgaps are reproduced, which closely match band gap values deduced from experiment \cite{Umari2012,Xia2013}.  This approach allows us to obtain more accurate orbitals than found via the PBE functional. In particular, this choice better reproduces the band dispersion of the unoccupied bands which affects the localization and hence the binding energy of the exciton. As will be shown in Fig.\ \ref{fig:allbse}, using PBE orbitals and energies misestimates the band dispersion and hence the localization of the exciton. This results in underestimated amplitude ratios and unsatisfactory relative peak positions. Conduction band DOS and band structures are also computed with PBE0($\alpha=0.24$). Note that in Ref.\ \cite{Kramberger2007} PBE was used to calculate the conduction band DOS. 

The method based on the final-state approximation used in this study is the full core-hole (FCH) method. In the FCH method, the excited electron is not added to the lowest conduction band but to the background charge \cite{Michelitsch2019}. It was shown that  X-ray absorption spectra are improved with this procedure \cite{Unzog2022}. To calculate FCH spectra we use an implementation presented previously \cite{Karsai2018}. The electronic states were determined using PBE0 and we denote these spectra as FCH. The spectra converged with a supercell size of 7 nm along the tube axis and for all FCH spectra only the $\Gamma$ point was sampled.

To calculate $GW$+BSE spectra we use the core-hole BSE methodology as implemented by Unzog \emph{et al.}\ \cite{Unzog2022}. To determine the screened interaction, we made use of the low-scaling $GW$ algorithm of  Kaltak \emph{et al.}\ \cite{Kaltak2014}. As it accurately describes the screening and yields good bandgaps within $G_0W_0$, we use the RPA based on PBE for the screening in the BSE which we denote as  $W_0^{\text{PBE}}$. In the set-up of the Bethe-Salpeter eigenvalue equation the PBE0($\alpha$=0.24) ground-state energies and orbitals are used. In the following, we call these spectra BSE for short. When BSE calculations are performed on top of $G_0W_0$@PBE, the intensity of the first peak is way underestimated to the rest. Using PBE0 orbitals and energies yields qualitatively better results. We believe PBE0 orbitals are better approximations for the Dyson orbitals, but orbital selfconsistent $GW$ calculations are prohibitive for the systems considered here.  In Tab.\ \ref{tab:bseparam} we list the computational parameters used for the BSE spectra. 

As usual, we made the static approximation for the screened interaction when solving the BSE. A soft  projector augmented-wave (PAW) \cite{bloechl1994} pseudopotential was used for carbon (``C\_s\_GW") and the plane-wave cutoff for orbitals and the cutoff of the response function were set to 300 eV and 150 eV, respectively. We used the Tamm-Dancoff approximation \cite{Dancoff1950,Tamm1991} for all BSE spectra and excluded excitations from valence to conduction band states. We calculated the BSE dielectric function in the $\mathbf{q}\to 0$ limit, see Eq.\ (29) of \cite{Unzog2022}. 

We point out that the EELS experiments of Senga \emph{et al.}\ \cite{Senga2016} involve a finite momentum transfer due to the large convergence and collection angles of  42.5 mrad and 33 mrad, respectively. However, in the valence-loss spectrum of Fig.\ 3b of Ref.\ \cite{Senga2016} we see mostly dipole-allowed $E_{ii}$ transitions while  additional multipolar contributions such as intraband excitations or activation of dark excitons \cite{Senga2018} are hardly visible. Therefore, the EELS setup of Ref.\ \cite{Senga2016} is dipole-dominated which allows us to directly compare the FCH and BSE results with the EELS spectra.

Within the PAW approach implemented in VASP the core states are frozen \cite{bloechl1994,kresse1996}, and absolute core energies are not computed. Instead, the spectra are shifted by aligning the first peak of the modeled spectra with the first peak observed in experiments. The modeled spectra are obtained by averaging the three diagonal components of the imaginary dielectric tensor, and a Lorentzian broadening of 0.2 eV (full width at half maximum) was applied to all modeled spectra.

\section{Results}\label{sec:results}

\subsection{Spectra}\label{sec:spectra}

\begin{figure}
    \centering
    \includegraphics{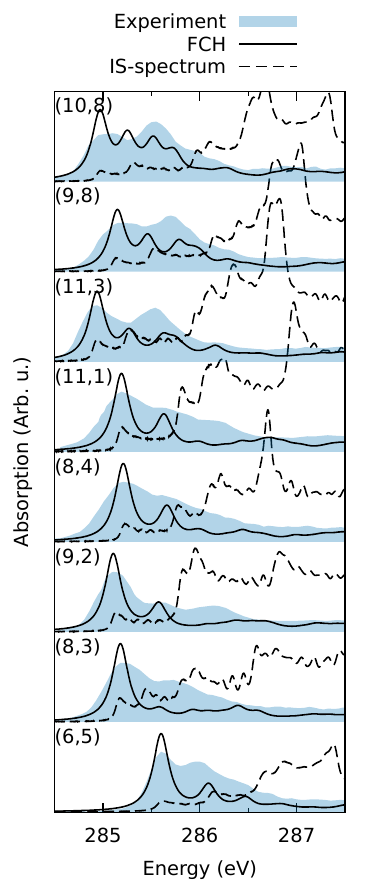}
    \caption{Comparison of experimental and modeled \emph{K}-edge absorption spectra of isolated SWCNTs. \emph{Red lines:} Experiments by Senga \emph{et al.}\ \cite{Senga2016}.  \emph{Black solid lines:} FCH spectra using PBE0, see Sec.\ \ref{sec:methods} for details. \emph{Black dashed lines:} Initial state spectra, \emph{i.e.}\ the PBE0 DOS where each conduction band energy is weighted by the transition probability from the core state to the corresponding conduction band state. The SWCNTs are arranged so that the radius decreases from top to bottom. All FCH spectra are shifted to align the first peaks in the experimental and modeled spectrum and scaled such that the area below the first peak of modeled and experimental spectra are approximately equal.} 
    \label{fig:fch}
\end{figure}

In this section, we compare our results to the experiments and relate the VHSs to the peaks at the absorption edge. Kramberger \emph{et al.}\ \cite{Kramberger2007} used the conduction band DOS to compare the positions of the VHSs with the peaks at the absorption edge. We show instead for each SWCNT the initial state (IS) spectrum. The IS spectrum is obtained from the conduction band DOS in the absence of the hole by weighing each conduction band energy by the dipole transition matrix element from the core state to the corresponding conduction band state.  It is important to note that this method is equivalent to the independent particle (IP) approximation, a common term in the many-body community, and we will use both terms (IP and IS) interchangeably.
To find the correspondence between the VHSs and the peaks in the IP spectrum we followed Kramberger \emph{et al.}\ \cite{Kramberger2007} and shifted each IP spectrum so that the first VHS aligns with the first peak at the absorption edge.  

The spectra of individual SWCNTs are shown in Fig.\ \ref{fig:fch}. Since the computational complexity of the FCH method is that of a DFT calculation we used it to model spectra of all semiconducting SWCNTs investigated in Ref.\ \cite{Senga2016}. In (6,5), (8,3), (9,2), (8,4), and (11,1)  FCH can resolve the most pronounced peaks of the fine structure visible in the experimental absorption edges. The relative positions of the second peaks are generally in better agreement with experiment than the third and fourth peaks. In the measured spectra of (9,8), (10,8), and (11,3) we can see two broad peaks in the absorption edge which we interpret individually as two close by peaks. The FCH spectra can tentatively replicate these features although the first two peaks are slightly too far apart. The energy separations of the third and fourth peaks in the FCH spectra more closely match the corresponding broad second peak in experiment. The broad peak formed by the third and fourth peaks aligns with the corresponding peak for the (10,8) spectrum while it is blueshifted in (9,8) and (11,3). 

Comparing IS and FCH spectra, we can see that the positions of the VHSs are significantly shifted from the peaks of the fine structure. If core-hole effects were negligible, then the relative positions of the VHSs should closely match the energies of the peaks at the absorption edge. The large mismatch between absorption peaks and VHSs in the IS spectra shows instead that core-hole effects significantly alter the conduction band energies. A systematic shift of the VHSs relative to the absorption peaks as a function of radius is not visible from Fig.\ \ref{fig:fch}.

\begin{figure*}
    \centering
    \includegraphics{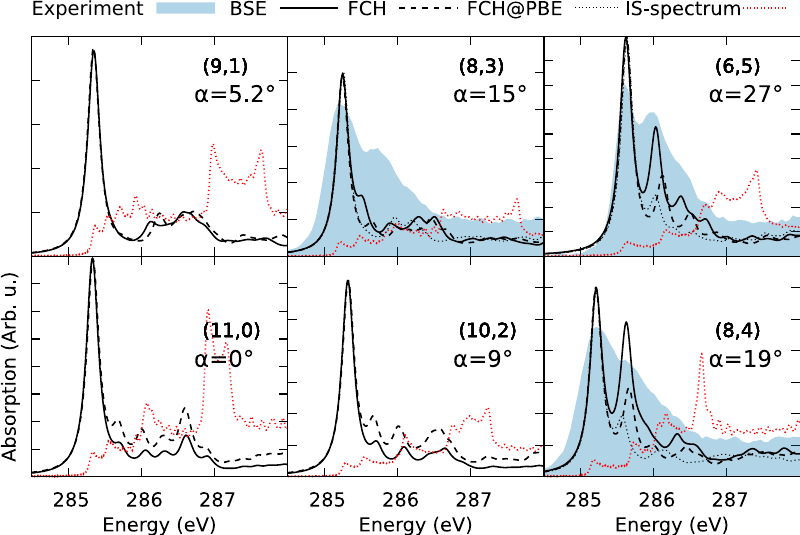}
    \caption{Comparison of modeled \emph{K}-edge spectra of isolated SWCNT with their initial state spectra. \emph{Black solid lines:} BSE spectra using PBE0 orbitals and energies and $W_0^{\text{PBE}}$ in the BSE Hamiltonian. \emph{Black dashed lines:} FCH spectra using PBE0. The BSE and FCH methods are explained in Sec.\ \ref{sec:methods}. \emph{Red dotted lines:} Initial state spectra. \emph{Blue filled areas:} Experimental spectra of Senga \emph{et al.}\ \cite{Senga2016}. All modeled spectra are shifted to align the first peaks in experimental and modeled spectra. The modeled spectra are scaled such that the area below the first peak of the modeled and experimental spectra are approximately equal.}
    \label{fig:allbse}
\end{figure*}

We proceed with the BSE results. The high computational requirements of the BSE approach prohibit us from calculating spectra of all SWCNTs investigated in Ref.\ \cite{Senga2016}. Therefore, we present in Fig.\ \ref{fig:allbse} BSE and FCH spectra for a subset of chiralities. The average radius of the SWCNTs in the top (bottom) row is 3.8 Å (4.3 Å). In each figure, we indicate the chiral angle of the SWCNTs \cite{Saito1998}. For chiralities (9,1), (11,0), and (10,2) we have no experimental data at our disposal. The BSE spectra give slightly better relative peak positions for (8,4) and (6,5). This is particularly evident for the chirality (6,5), where BSE accurately matches the positions of all four sub-peaks in the absorption edge, whereas these features are blueshifted with FCH. Additionally, the amplitude of the second peak in (6,5) and (8,4) is higher in BSE compared to FCH. The improvement of BSE over FCH aligns with the findings of Unzog \emph{et al.}\ \cite{Unzog2022} for reference bulk systems. Although $G_0W_0$ on top of PBE rectifies the peak positions, it does not change the intensity distribution between the peaks. This issue is also visible in BSE calculations, whenever PBE orbitals are used, and motivates our choice to start BSE calculations from PBE0 orbitals and eigenvalues. 

Our choice to describe the orbitals using hybrid functionals is motivated by comparing the FCH spectra obtained with PBE and PBE0 functionals shown in Fig  \ref{fig:allbse}. While the spectrum based on  PBE orbitals shows an overall similar peak structure, some of the secondary peaks are shifted and the amplitudes are underestimated. 

Similarly to the FCH results, comparing the peaks in the BSE spectra to the VHSs in the IS spectra reveals large discrepancies for all systems. A particularly large difference can be seen for  (9,1) where the IS spectrum shows four VHS in the range 285 eV to 286 eV. Both modeled spectra of (9,1) show instead only a single broad peak. By inspecting Fig.\ \ref{fig:allbse} we could infer whether the shift of a peak at the absorption edge with respect to the VHS is correlated with the chiral angle. As with the curvature above, we do not observe a systematic correlation. 

In discussing optical spectroscopies of SWCNTs they are separated into two families depending on whether $(n-m) \mod 3$ is equal to 1 or 2 \cite{Sfeir2006,Choi2013}. By inspecting Figs.\ \ref{fig:fch} and \ref{fig:allbse}, we can also exclude that misalignment between peaks at the absorption edge and VHSs is only found in one of those families or that one family shows more misalignment than the other. 

\raggedbottom

\subsection{(6,5) fatband structure}\label{sec:fatband}

These findings indicate that the relationship between VHSs and the peaks at the absorption edge is more complicated than suggested by Kramberger \emph{et al}.\ \cite{Kramberger2007}. In that work each peak in the fine structure is identified with one particular VHS. To understand how to relate the VHSs with the peaks at the absorption edge we investigate the fatband structure \cite{Bokdam2016} of the (6,5) SWCNT. In the BSE formalism, the exciton is formed by all possible electron-hole pairs. The so-called BSE coupling coefficients quantify how much a particular electron-hole pair contributes to the exciton. Since the core state is fixed in our calculations we can use the BSE coefficients to infer which conduction band states contribute more strongly. In the fatband structure, the squared magnitudes of the BSE coupling coefficients of a state within a given band and at a given $\mathbf{k}$-point are drawn as circles superimposed on the band structure. 

We point out differences of about 50-100 meV between the centers of some circles, \emph{i.e.}, the conduction band energies used in the BSE Hamiltonian, and the corresponding states in the band structure. We attribute these differences to the separate treatment of the $|\mathbf{q}+\mathbf{G}|=0$ singularity of the Coulomb interaction in reciprocal space $V=4\pi/|\mathbf{q}+\mathbf{G}|^2$, where $\mathbf{G}$ are the reciprocal lattice vectors and $\mathbf{q}=\mathbf{k}-\mathbf{k}'$. For the BSE spectra, the singularity correction of Massidda, Posternak, and Baldereschi \cite{Massidda1993} was used in two cases: in the determination of the PBE0 ground-state orbitals and energies and when solving the BSE. In contrast, when calculating the band structures, the Coulomb interaction was truncated according to Spencer and Alavi \cite{Spencer2008}, as only truncated Coulomb kernels allow for an interpolation to dense $\mathbf{k}$-point grids.

\begin{figure*}
    \centering
    \includegraphics{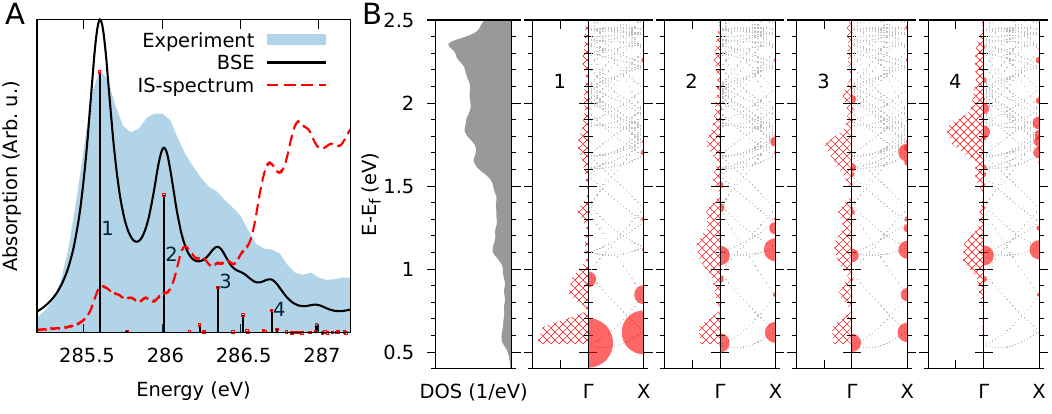}
    \caption{The fatband structure of the (6,5) absorption edge. Panel \textbf{A}: Enlarged view of the absorption edge. \emph{Blue filled area:} Experiment by Senga \emph{et al.}\ \cite{Senga2016}. \emph{Black solid line:} BSE spectra using PBE0 orbitals and energies and $W_0^{\text{PBE}}$ in the BSE Hamiltonian, see Sec.\ \ref{sec:methods}. The BSE excitations are shown as black lines. The first four bright transitions are labeled 1-4. \emph{Red dashed line:} Initial state spectrum. Panel \textbf{B}: Fatband structures of the first four bright excitations of panel A. \emph{Gray shaded area:} Initial state spectrum. \emph{Panels 1-4:} BSE fatband structures of transitions 1-4 of panel A. The PBE0 band structure is shown with gray dots. The radii of the red circles are proportional to the square of the BSE coefficients. We show histograms of the fatband structures with red-shaded areas.  Only contributions at $\Gamma$ and $X$ are shown, as $\mathbf{k}$-point sampling in the BSE was limited to these two $\mathbf{k}$-points.} 
    \label{fig:65fatband}
\end{figure*}

We present in panel A of Fig.\ \ref{fig:65fatband} an enlarged view of the absorption edge. We have marked with labels 1-4 the excitations corresponding to the four brightest absorption peaks visible in the fine structure.  We can see a small shift of the second VHS in the IS spectrum  to the second excitation (labeled 2) in the BSE spectrum, while the third and fourth VHSs are significantly blue-shifted with respect to the corresponding BSE excitations (3 and 4).

The fatband structures of the first four bright transitions 1-4 shown in panel A are shown in panel B. For the excitation 1,  the bands of the first VHS at around 0.75 eV contribute the most. This contribution of the first VHS is in line with a simple correspondence between a peak in the fine structure and a particular VHS. We can also see some contributions of higher-lying conduction band states which decrease in strength with increasing energy. 

We also see a very sizeable contribution from the edge of the Brillouin zone that does not fit with the simple picture of the exciton being related to a single VHS. The exciton is rather formed by linearly combining low-energy states with progressively less contributions from higher energy states. This is exactly what one would expect from a local exciton build from upwards dispersing $\pi^*$-bands. The specifics of the band structure and Brillouin zone do not matter. This will also be later confirmed by the exciton wavefunction analysis.

Investigating the fatband structure of excitation 2 we can see that the bands of the second VHS located at 1.1 eV are strongly involved in forming the exciton. However, we can also see additional contributions of conduction band states from the first VHS. The histogram further shows that the contributions from the first and second VHS are approximately equal in magnitude. This pattern holds for the excitations 3 and 4 as well. Because of the mutual orthogonality of the wavefunctions of different excitations, this pattern is a direct consequence of the first exciton involving orbitals from higher energies. Likewise, since the second exciton involves orbitals at even higher energies, for those to then be orthogonal to the first and second exciton, they also need to involve contributions from orbitals related to the first and second VHS. This means that the excitations generally involve conduction band states across multiple VHSs. Therefore, we have to consider that a wide range of transitions into different VHSs are coupled to form one exciton. It is therefore difficult to predict peak positions at the absorption edge only from the relative distances of VHSs. On the contrary, a perfect match between VHSs and a peak at the absorption edge is coincidental because the excitonic energy results in a complex manner from the transitions to different VHSs.

It could be argued that in cases where close alignment between a VHS and a peak in the fine structure is found, only states from that particular VHS are involved. We, however, emphasize that we need to conclude that multiple VHSs contribute, although the energy difference between the first and second peak seemingly often matches the energy difference between the first and the second VHS. Even in this case, conduction band states across multiple VHSs contribute to the exciton.

\begin{figure}
    \centering
    \includegraphics[width=0.9\linewidth]{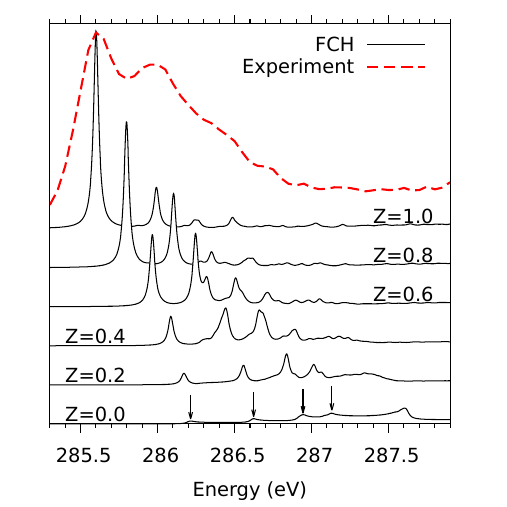}
    \caption{Increasing the core-hole charge in FCH spectra of chirality (6,5).  \emph{Black lines:} FCH spectra using PBE, with core-hole charge indicated. \emph{Red line:} Experiment \cite{Senga2016}. All spectra are shifted and scaled equally, with the shift chosen to align the first peak of the $Z=1$ spectrum with the first peak in experiment. Arrows mark the first four VHSs in the $Z=0$ spectrum.}
    \label{fig:65_adia}
\end{figure}

To investigate how core-hole effects change the VHSs we show in Fig.\ \ref{fig:65_adia} FCH spectra for  chirality (6,5), where the core-hole charge $Z$ is incrementally increased from 0 to 1 atomic unit. For this result, we use FCH based on PBE orbitals. By comparing the spectra with $Z=1$ and $Z=0$ we conclude firstly that the first and second exciton have binding energies of 0.6 eV and 0.2 eV, respectively. Second, we again conclude that core-hole effects are significant, as the downward shift is comparable in magnitude to the separation between adjacent VHSs. Third, the shifts of the VHS are not uniform as different VHSs are shifted to varying degrees. Finally, large shifts of oscillator strength into the first four VHS are visible upon increasing $Z$. These shifts make it difficult to determine which states in the $Z=0$ spectrum contribute to a particular absorption peak in the $Z=1$ spectrum. This further supports the conclusion above that a wide range of transitions form one exciton. We note that the $\pi^*$ response around 287.6 eV disappears when the core-hole charge is switched on. It is tempting to attribute the increase in oscillator strength of the first four VHSs to states from the $\pi^*$ peak. However, when we re-examine the fatband structure in Fig.\ \ref{fig:65fatband} we see that none of the first four absorption peaks have significant contributions from the corresponding VHS here located at 2.5 eV in the IS spectrum.

\raggedbottom

\subsection{Graphene supercell}\label{sec:periodicity}

\begin{figure}
    \centering
    \includegraphics[width=\linewidth]{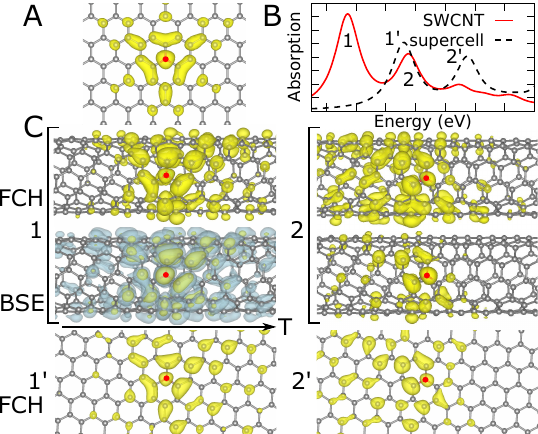}
    \caption{Comparison of spectra and wavefunctions of a (6,5) SWCNT and a corresponding graphene supercell. Panel \textbf{A}: Charge density of the exciton in graphene. Panel \textbf{B}: BSE spectra of the SWCNT (\emph{red solid line}) and the equivalent graphene supercell (\emph{black dashed line}).  Panel \textbf{C}: Charge densities of the excitons corresponding to excitations 1 and 2 in panel B using FCH (top row) and BSE (middle row). The excitons of transitions 1' and 2' of panel B (supercell) are shown in the bottom row. The core-excited atom is marked with a red  dot (black in greyscale) and the value of the yellow isosurface is set to $5\times10^{-4}e$ throughout. The blue isosurface is set to $1\times10^{-4}e$. The direction of the translation vector $\mathbf{T}$ of the supercell is indicated in the figure. The charge density plots were generated with VESTA \cite{Momma2011}.}
    \label{fig:graphene_results}
\end{figure}

We now investigate why conduction band states across multiple VHSs are involved in creating a particular excitonic transition. To eliminate curvature effects we map the unit cell of the SWCNT onto a graphene supercell, where the basis vectors are the chiral and translation vectors  $\mathbf{C}$ and $\mathbf{T}$, respectively, and $\mathbf{k}=k_c\mathbf{K_1} + k_t \mathbf{K_2}$, where $\mathbf{K_1, K_2}$ are the reciprocal lattice vectors along the circumference and nanotube axis, respectively \cite{Saito1998}. Along the tube axis $k_t$ is sampled on a fine grid, whereas along the circumference $k_c=0$. This way, the boundary conditions of the SWCNT are reproduced, \emph{i.e.}, infinite extent along the nanotube axis and confinement along the circumference with an added periodic boundary condition. The results below were obtained for a graphene supercell corresponding to the chirality (6,5). When both $k_c$ and $k_t$ are finely sampled we recover the \emph{K}-edge spectrum of graphene. In contrast to the SWCNTs, the $\pi^*$ response of graphene is dominated by a single excitonic transition \cite{Hua2010}. We display the charge density of this exciton in panel A of Fig.\ \ref{fig:graphene_results}. 

The spectra and charge densities of the excitons are collected in Fig.\ \ref{fig:graphene_results}. In panel B we show the BSE spectra of the (6,5) SWCNT and supercell with red and black dashed lines, respectively. Both spectra are multiplied by the volume of the cell and shifted by the energy of the conduction band edge. We can see that the supercell spectrum also exhibits the characteristic fine structure observed in the absorption edge of the SWCNT spectrum. The supercell spectrum is however blueshifted with respect to the SWCNT spectrum and the amplitude of the first peak is decreased. We attribute these effects to the curvature of the SWCNT which reduces the dispersion of the $\pi^*$ bands \cite{Blase1994} and in turn increases the binding energy of the exciton. We mark the first two excitations of the SWCNT and supercell spectra with \{1,2\} and \{1',2'\}, respectively. 

In panel C, we show the charge densities of the first two bright excitations in the (6,5) SWCNT and in the supercell. The hole was fixed at the excited atom highlighted with a red dot (black in greyscale). The charge densities corresponding to the first (1 and 1') and second (2 and 2') bright peaks are presented in the first and second columns, respectively. For the SWCNT, we show the charge densities obtained from FCH and BSE methods.

We first investigate the charge densities in the first column corresponding to the peaks 1 and 1' of panel B.  We can see that the excitons corresponding to peaks 1 and 1' closely resemble the exciton of graphene shown in panel A. In particular, all three excitons demonstrate approximately the same symmetry and localization. The BSE charge density is more localized than the FCH charge density, though.  To facilitate the comparison between FCH and BSE we provide a second isosurface with a smaller value which emphasizes the symmetry better.

While the first exciton resembles the exciton in graphene, excitons 2 and 2' are much more extended. Curvature effects are absent in the graphene supercell model. Therefore, we attribute this extended shape entirely to the $\mathbf{k}$-point sampling in the supercell, which reproduces the boundary conditions of the SWCNT.

The difference between the first and second excitons can be explained as follows. We determine the bound states of the core-hole potential while considering the boundary conditions in the SWCNT. The core-hole potential is strong enough that the first bound state has a large binding energy. It is sufficiently strongly bound that the first exciton is not affected by the confinement along the circumference and hence completely localized near the core-hole. Its shape is then determined by a few nearest neighbors of the excited atom so that the exciton resembles the exciton in graphene. In contrast, the second exciton has a reduced binding energy and a more delocalized wavefunction. For the second exciton, the  confinement along the circumference must be considered, resulting in a wavefunction that is not determined by the nearest neighbors but exhibits the extended shape visible in Fig.\ \ref{fig:graphene_results}.

\begin{figure}
    \centering
    \includegraphics[width=0.9\linewidth]{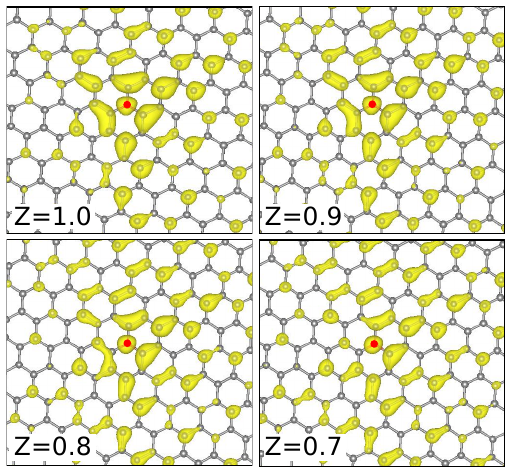}
    \caption{Charge density of the lowest energy exciton in a graphene supercell that reproduces the boundary conditions of a (6,5) SWCNT. The value of the core-hole charge $Z$ is indicated in each panel. The charge density plots have been obtained using FCH based on PBE orbitals, see text for details. The excited atom is marked with a red dot (black in greyscale) and the charge density plots were generated with VESTA \cite{Momma2011}.}
    \label{fig:wf_adia}
\end{figure}

In short, strongly bound excitons resemble the exciton in graphene, while reducing the binding energy of an exciton results in a shape that adapts to the periodic boundary conditions. To test this idea we use the FCH method which allows us to adjust the core-hole charge $Z$ and hence the binding energy $\sim Z^2$. We show in Fig.\ \ref{fig:wf_adia} the charge densities of the first bound state in the graphene supercell using PBE orbitals and different values for $Z$. The excited atom is marked with a red dot (black in greyscale). The value $Z=0.7$ leads to a state with about half the binding energy of the $Z=1$ state. The continuous change from a shape resembling the exciton in graphene to an extended shape is evident.

\begin{figure}
    \centering
    \includegraphics[width=0.9\linewidth]{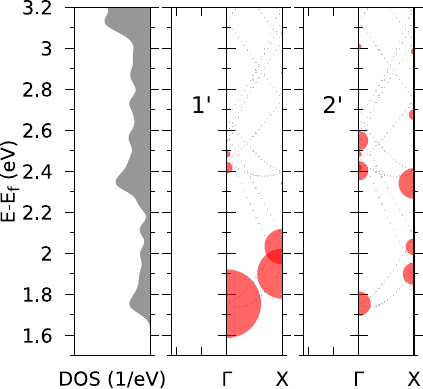}
    \caption{The BSE fatband structures of the first two bright transitions (1' and 2') of the graphene supercell spectrum shown in panel A of Fig.\ \ref{fig:graphene_results}. The radii of the red circles are proportional to the amplitude square of the BSE coefficients. The initial state spectrum is shown with gray dots.}
    \label{fig:graphene_fatband}
\end{figure}

To relate these results to the previous section we plot in Fig.\ \ref{fig:graphene_fatband} the fatband structures of the first two excitations of the graphene supercell, \emph{i.e.}, excitations 1' and 2' in panel B of Fig.\ \ref{fig:graphene_results}. Comparing Fig.\ \ref{fig:graphene_fatband} with Fig.\ \ref{fig:65fatband} we see the same structure: the first excitation couples bands mostly from the first VHS, while the second excitation couples bands from many VHSs. Therefore, we can attribute the fatband structures 2, 3, and 4 of Fig.\ \ref{fig:65fatband} to a failure of a simple model of a localized exciton resembling the exciton in graphene. In these cases, conduction bands from many different VHSs contribute to form the extended excitons.

\subsection{Exciton character}\label{sec:char}

In this section we investigate the character of the excitons in more detail. Excitons are classified into Frenkel \cite{Frenkel1931,Frenkel1931a} or Wannier-Mott \cite{Wannier1937,Mott1938} excitons, the former corresponding to the atomic limit where electron and hole are found on the same atom and the latter to large separations between electron and hole. An intermediate model is the charge-transfer (CT) exciton \cite{Knox1963} that is encountered in molecular \cite{Knupfer2004,Zhu2009,Cudazzo2015} or ionic crystals \cite{Overhauser1956,Knox1959,Zhang1994,Abbamonte2008}. An exciton is identified as a CT-exciton when the electron and hole are localized at different molecules or molecular subunits \cite{Knox1963,Wright1995}.  

We quantified the electron localization by determining the dipole moments $\mathbf{p}=\sum_{i\in V} \rho(\mathbf{r}_i) (\mathbf{r}_i -\mathbf{r}_{\text{exc}})$ of bright excitons in semiconducting chiral and zig-zag SWCNTs, where the excited atom was the reference point $\mathbf{r}_{\text{exc}}$. The analysis shows that all first bright excitons have a vanishing dipole moment, and additionally, the charge density is maximal at the excited atom. It is therefore fairly evident that the first excitons have Frenkel character. 

In contrast, higher bright excitons have a finite dipole moment of about 1-2 eÅ. The component of the dipole moments along the nanotube axis is in all cases much larger than the component perpendicular to it (pointing upward in Fig.\ \ref{fig:graphene_results}). This suggests that these excitons have a CT character, implying that the center of the electronic charge is to some extent shifted to neighboring atoms. However, it was shown previously that a non-vanishing dipole moment is not sufficient to assign CT character \cite{Petelenz2013}. For example, in molecular crystals, the localization of the charge distribution can be changed by shifting the hole which results in zero net dipole moment on average\cite{Cudazzo2015}.
To test this,  we excited the hole at the nearest neighbor along the tube axis of the marked atom in Fig.\ \ref{fig:graphene_results}.
We found that the dipole moment then points to the right, while it originally pointed towards the left. This suggests that the net polarization is zero in an experiment where multiple neighboring atoms are excited. Since the dipole moment is zero on average, the core excitons of higher excitations in SWCNTs are better described as charge resonance states \cite{Petelenz2013}. 

\section{Discussion}\label{sec:discussion}

In Sec.\ \ref{sec:results},  we inferred that core-hole effects are strong since they significantly alter the spectrum from the initial state (IS) approximation. Nevertheless, it is hard to deny that the VHSs in the IS spectra often seem to align well with the position of the quasiparticle (QP) peaks in the experimental spectrum. Here, we argue that the separation between VHSs in the IS spectrum determines whether the IS spectrum is a good approximation to the  QP spectrum. Our argument relies on the orthogonality of excitons and is complemented by a formal approach based on pseudopotential theory.

We begin with the $K$-edge of graphene, where a {\em single} bound exciton dominates the $\pi^*$ response \cite{Hua2010}. The remaining $\pi^*$ part of the spectrum is featureless and
roughly forms a parabolic band. In SWCNT, the first exciton closely resembles the first exciton of graphite, as shown above (see Fig. \ref{fig:graphene_results}). 
As the first exciton is localized in real space, it is necessarily a linear combination of $\pi^*$  conduction band states. States from the bottom of the conduction band dominate the exciton, but states from higher energies also contribute, as shown in Fig.\ \ref{fig:65fatband}: as one moves further upwards from the bottom of the conduction band, the contributions of the conduction band states to the first exciton decay with the energy separation. 

We have seen above that the second and third peaks in the excitonic spectrum contain significant contributions from the original second and third VHS, but they also contain some contributions from the 1st VHS. This is a consequence of the orthogonality condition. As the first exciton contains contributions from the entire $\pi^*$ band (including contributions from the 2nd and 3rd VHS), mutual orthogonality requires that states at higher energies also contain some admixture from the 1st VHS.

From this picture, the following conclusion emerges. If the energy separation between the VHS is large, then the 1st bound exciton will be predominantly formed from conduction band states close to the first VHS, with little contributions from higher VHS. The second peak in the spectrum, which needs to be orthogonal to the 1st bound exciton (and thus the 1st VHS), will now be dominated by states from the second VHS, and so on. Hence, in this case, the IS picture is at least qualitatively a good approximation to the spectrum.

Conversely, if the energy separation between the VHSs is small, the first exciton will have strong contributions from the second VHS and potentially even further VHS. As orthogonality to the first exciton must be maintained, states from the first VHS will contribute significantly to the second as well as other exciton. In this case, the IS picture will not necessarily correspond well with the QP spectrum. 

To make this more explicit, we briefly relate these observations to the orthogonalized plane-wave (OPW) method \cite{Herring1940} as formulated by Phillips and Kleinman \cite{Phillips1959}. We summarize the important points and adapt the notation of Ziman\ \cite{Ziman1972}. In the OPW method, an ansatz is made for the conduction band states $|\psi\rangle$ by orthogonalizing a linear combination of plane-waves $|\phi\rangle$ to the core-states $|b_t\rangle$:
\begin{equation}\label{equ:proj}
    |\psi\rangle = \left[1-\sum_t |b_t\rangle\langle b_t|\right] |\phi\rangle.
\end{equation}
This procedure results in an effective Hamiltonian (in atomic units) for the pseudo-orbital $\phi$
\begin{equation}\label{equ:ham}
    H_{\text{eff}}(E)=-\frac{1}{2}\Delta+\underbrace{V+\sum_t (E-E_t)|b_t\rangle\langle b_t|}_{\Gamma(E)},
\end{equation}
where $V$ is the attractive potential of the nuclei \cite{Phillips1959,Ziman1972}. The Hamiltonian is now expressed in terms of the effective pseudopotential $\Gamma(E) = V+V_R$, where $V_R=\sum_t (E-E_t)|b_t\rangle\langle b_t|$. As the conduction band energies $E$ are always larger than the core state energies $E_t$, $V_R$ is a positive, \emph{i.e.}, a repulsive potential that compensates the attractive potential of the nuclei $V$ \cite{Phillips1959}.

We can readily adapt the formalism to $K$-edge excitations. The carbon $1s$ hole is localized at the nucleus, and can be approximated by a positive point charge $V(+1)$ located at one of the nuclei. The potential $V(+1)$ leads to one excitonic bound state with a negative charge, and we can mimic the combined potential of the positive core charge
and the bound exciton in exactly the same manner as in the OPW theory, Eq. (\ref{equ:ham}), but with a single bound state $t$:
\begin{equation}\label{equ:newham}
H_{\text{eff}}=-\frac{1}{2}\Delta+V(+1)+(E-E_t)|b_t\rangle\langle b_t|,
\end{equation}
where $E_t$ is the energy of the bound exciton.

The effective potential felt by electrons at higher energies is then the sum of the original attractive V(+1) potential plus the repulsive potential caused by the requirement of orthogonalization.
As discussed before, if the bound exciton is predominantly 
made up of states in the first VHS, the overall potential experienced by the ``remaining" electrons is weak. Hence, the rest of the spectrum is dominated by the effects imposed by the quantization of electronic states in the circumferential direction, and concomitantly, the second peak in the spectrum is dominated by states from the 2nd VHS, etc. 

We note that this concept, matter of fact, shows again that the 2nd peak will always contain contributions from the 1st VHS, as the wave function $\psi$, by construction, possesses contributions from the first bound exciton due to the projection in Eq. (\ref{equ:proj}). From experience, it is known that the OPW concept is accurate in the limit that the bound state is localized and much lower in energy than the remaining states. Hence, if the energy separation between the VHS is large, and the bound exciton is dominated by a linear combination of the 1st VHS only, the core hole is mostly screened ($\Gamma(E)\approx 0$), and the remaining QP states resemble the IP spectrum.

\begin{figure}
    \centering
    \includegraphics[width=0.9\linewidth]{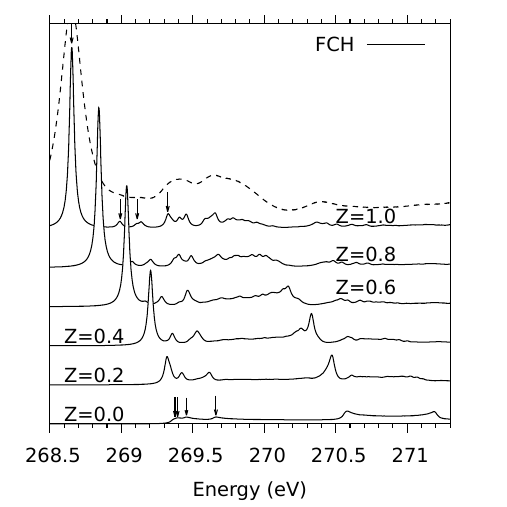}
    \caption{Increasing the core-hole charge in FCH spectra of chirality (9,1).  \emph{Black lines:} FCH spectra using PBE, with core-hole charge indicated. \emph{Black dashed line:} $Z=1$ spectrum using a broadening 0.2 eV (full-width half maximum) and scaled by a factor 6 for easier comparison with Fig.\ \ref{fig:allbse}. Arrows mark the first four VHSs in the $Z=0$ spectrum and their corresponding peaks in the $Z=1$ spectrum.}
    \label{fig:91_adia}
\end{figure}

Summarizing, we argue that the energy difference between VHSs determines whether an approach based on VHSs in the IP spectra leads to satisfactory predictions for the QP peaks. This is also illustrated in Fig.\ \ref{fig:allbse}. For the (6,5) chirality, the first three VHSs are widely spaced, resulting in absorption spectra that qualitatively match the IS spectra. Conversely, for the (9,1) chirality, the first four VHSs are close together, and the actual QP absorption spectrum does not quite resemble the IS spectrum. To make this failure for (9,1) more apparent, we show in Fig.\ \ref{fig:91_adia} spectra with an increasing core-hole charge
(for the (6,5) tube the corresponding results are shown in Fig. \ref{fig:65_adia}). The first four VHSs in the IS spectrum and their corresponding peaks in the FCH spectrum are marked with arrows, highlighting the large discrepancies: upon increasing the core-hole charge $Z$, the 
local exciton is formed as a linear combination of the conduction band $\pi^*$ orbitals.
The exciton separates from the remaining peaks and shifts down in energy. Remarkably even in this case, the separation between the 2nd, 3rd and 4th peaks in the final QP spectrum is quite similar to the separation between the corresponding three peaks in IS sp the IP approximation ($Z=0.0$). However, the final  spectrum also shows additional fine structure beyond the 4th peak that is absent in the IP picture. 

Within this framework, we can also understand the results on $C_{60}$ fullerenes where the absorption spectrum closely resembles the IS spectrum \cite{Nyberg1999}. The VHSs in the IS spectrum are at least 1 eV apart, almost twice the value of the separation between the first three VHSs in (6,5) where the qualitative picture was shown to hold. In this case, the gaps between the VHSs are so large that hardly any states of the second VHS contribute to the first exciton. In other words, the orthogonalization contribution $(E-E_t)|b_t\rangle\langle b_t|$ can largely compensate for the attractive core-hole potential.

We emphasise, however, that even if the energy separation between the VHSs is large, the IS spectra are at best qualitatively correct. For a quantitative comparison, core-hole effects should be included, since the peaks visible in the spectrum are always linear combinations of many states of at least one, but often several, VHSs. Furthermore, the exact position of the resonances depends on a balance between the effects caused by the partially shielded core hole and the quantisation of the electronic states around the tube-circumference.

%------------------------------------------------------------------------------
\section{Conclusion}\label{sec:conclusion}
%------------------------------------------------------------------------------

In this paper, we have modeled X-ray spectra of isolated SWCNTs and compared them with experimental EELS spectra \cite{Senga2016}. We used the supercell approach in a version we call the Full Core Hole (FCH) method, as well as the Bethe-Salpether approach (BSE). In both methods, we used the global hybrid functional PBE0 either as a starting point for BSE or in the FCH calculation, since it provides improved orbitals and a better description of the band dispersion.

The FCH and the BSE spectra are in reasonable agreement with each other.
The spectra show also satisfactory agreement with the experimental data, with the BSE demonstrating slightly better alignment with the experimental results.
The core-hole effects are very significant both, in the FCH and the BSE method. Specifically, core-hole effects lead to a significant shift of the intensities from higher energies to low energy excitations, but they also change the energy difference between the peaks significantly. So the common approach to compare excitation energies in the independent particle picture with experimental excitation energies seems to be at least somewhat questionable. To investigate the excitonic effects in more detail,  we used a fatband analysis. This shows that conduction band states from multiple VHSs contribute to one exciton.

To investigate whether curvature effects are strong, we also performed graphene supercell calculations with a $\mathbf{k}$-point sampling chosen to mimic the quantization effects imposed on the electronic states around the circumference of the tube.
We found that the core-hole spectra are very similar for graphite and nanotubes. However, the curvature reduces the bandwidth, which in turn increases the excitonic effects in carbon nanotubes.
Overall, the spectra are nevertheless only slightly different for "quantized" graphite and nanotubes.

A key finding of the present study is that the first exciton is strongly localised in all nanotubes and has a similar character and electron charge density distribution as in graphene. It is best described as a Frenkel exciton. In graphene, however, no additional QP peaks are visible, neither experimentally nor theoretically. It is therefore clear that the additional peaks in the core-hole spectrum, must be a result of the effects imposed by the quantisation of the electronic states around the circumference of the tube. However, all our results suggest that there is no simple one-to-one correspondence between the independent particle spectrum and the full QP spectrum. For example and as mentioned above, a fatband analysis shows that the QP peaks contain orbital contributions from many VHS and the position of the peaks is different for the IP spectrum and the spectrum including core-hole effects. Although the details differ between the spectra with and without core-hole effects, we have been able to find a theoretical motivation why, for all but the first bound exciton, there is usually a strong similarity for the VHS and the higher energy peaks in the QP spectrum. This theoretical motivation is based on the orthogonalised plane wave method. As described above, the first exciton is bound and the requirement of orthogonalisation of the higher excitons to the first exciton implies that the core hole is  ``screened" at higher excitation energies by the first exciton. 
Thus, at higher excitation energies the positions of the QP peaks are often quite close to the positions of the VHS. Whether this picture is correct depends on whether the bound first exciton is predominantly composed of states in the first VHS or not.
If the VHS have a large energy difference, then there is often a close match between the positions of the VHS in the IP spectrum and the QP peaks in the full core-hole spectrum. However, for an accurate description, it seems expedient to perform full BSE calculations, ideally starting from PBE0 orbitals.

\section*{Acknowledgements}

This research was funded in whole or in part by the Austrian Science Fund (FWF) DOC 85-N. For open access purposes, the author has applied a CC BY public copyright license to any author-accepted manuscript version arising from this submission. This study is part of a project that has received funding from the European Research Council (ERC) under the European Union’s ERC-Synergy Project ``MORE-TEM" (951215). The computational results presented were achieved using the Vienna Scientific Cluster (VSC-5). We thank Christian Kramberger, Francesco Sottile, and Lorenzo Sponza for valuable feedback and discussion.

\raggedbottom

\bibliography{bibliography} 
\end{document}